\documentclass[aps,pre,twocolumn,reprint,longbibliography,titlepage,showpacs,showkeys,amsfonts,amssymb,amsmath,footinbib,raggedbottom,floatfix]{revtex4-1}
\usepackage{graphicx}
\usepackage{epstopdf}
\usepackage{color}
\usepackage{bm}
\usepackage{pdfsync}
\usepackage{url}
\newcommand{\bse}{\begin{subequations}}
\newcommand{\ese}{\end{subequations}}
\newcommand{\be}{\begin{equation}}
\newcommand{\ee}{\end{equation}}
\newcommand{\bea}{\begin{eqnarray}}
\newcommand{\eea}{\end{eqnarray}}
\newcommand{\kb}{k_{_{\mathrm{B}}}}

\newcommand{\tp}{t^{\prime}}

\newcommand{\phiv}{\varphi_{_{\!\mathbf{v}}}}
\newcommand{\phivzero}{\varphi_{_{0}}}
\newcommand{\chiq}{\chi_{_{\mathbf{q}}}\!}
\newcommand{\chiqp}{\dot{\chi}_{_{\mathbf{q}}}\!}
\newcommand{\chiv}{\chi_{_{\mathbf{v}}}\!}
\newcommand{\chivp}{\dot{\chi}_{_{\mathbf{v}}}\!}
\newcommand{\sigmaQ}{\sigma_{\mbox{\tiny{\!\bf{Q}}}}}
\newcommand{\sigmaQp}{\dot{\sigma}_{\mbox{\tiny{\!\bf{Q}}}}}
\newcommand{\DQ}{D_{_{\mbox{\tiny{\!\bf{Q}}}}}\!}
\newcommand{\sigmaone}{\sigma_{_{\!\mathbf{1}}}}
\newcommand{\Done}{D_{_{\mbox{\tiny{\!\bf{1}}}}}\!}

\newcommand{\sigmac}{\sigma_{\mbox{\tiny{\!\bf{clas}}}}}
\newcommand{\qzero}{q_{_{\mathbf{0}}}}
\newcommand{\vzero}{v_{_{\mathbf{0}}}}
\newcommand{\nun}{\nu_{_{\!\mathrm{n}}}}
\newcommand{\lambdaone}{\lambda_{_{\mathbf{1}}}}
\newcommand{\lambdatwo}{\lambda_{_{\mathbf{2}}}}
\newcommand{\csch}{\mbox{csch}}
\newcommand{\omegaD}{\omega_{_{\mbox{\tiny{\bf{D}}}}}}
\allowdisplaybreaks

\begin{document}
\title{Generalized diffusion of \textcolor{black}{quantum} Brownian motion}
\author{Pedro J. Colmenares}
\email{gochocol@gmail.com}
\affiliation{Departamento de Qu\'{\i}mica. Universidad de Los Andes. M\'erida 5101, Venezuela}
\thanks{Corresponding Author}


\begin{abstract}
This article discusses the numerical result predicted by the quantum Langevin equation of the \textcolor{black}{generalized} diffusion function of a Brownian particle immersed in an Ohmic quantum bath of harmonic oscillators. The time dependence of the standard deviation of  the reduced Wiener function of the system, \textcolor{black}{obtained by integrating the whole function in the momentum space,} is determined as well. The complexity of the equations leads to resort to a much simpler calculation based in the position correlation function. They are done for the three possible regimes of the system, namely, periodic, aperiodic and overdamped. It is found in the periodic case that the generalized diffusion \textcolor{black}{is a discontinuous function exhibiting negative values} during short time periods of time. This counterintuitive result, found theoretically in other systems and waiting for its experimental confirmation, can be perfectly explained in the framework of the quantum Langevin equation. \textcolor{black}{Its oscillatory behavior is primordially due to the response to the external field while its quantum origin contributes only in its magnitude}. The results are compared with those of the continuum limit which exhibits similar behavior. 
\end{abstract}

\maketitle
\vspace{-1.8cm}
\section{Introduction}

The Fokker-Planck equation (FPE) of a harmonically bound Brownian particle  has been extensively analyzed.  

Iit was originally studied \textcolor{black}{ in classical systems} by Chandrasekhar \cite{Chandrasekhar} and subsequently refined by Adelman \cite{Adelman1} and Adelman et al. \cite{AdelmanGarrison}. 

\textcolor{black}{The extension to a quantum harmonic thermal bath has also been developed by several authors.} For instance, Caldeira et al. \cite{CaldeiraLeggett} applied the  Feynman-Vernon theory \cite{FeynmanVernon} to study the quantum dissipation by assuming the initial preparation of the system has none effect on the evolution of the dynamics. They found a FPE by making a transformation from Hilbert space to the classical phase space by way of the Wigner distribution functions. An important advance in the search for an exact FPE was the work by Grabert el al. \cite{GrabertWeissTalkner} where the correlation functions of position and momentum were evaluate in closed form. Schramm et al. \cite{SchrammJungGrabert} partially used these results to enhance the FPE of Ref. \cite{CaldeiraLeggett} by expanding the time evolution of the probability density in terms of cumulants of the noise, previously presented by  H{\"a}nggi \cite{Hanggi1}, and including the correlation between the quantum noise and the particle initial position. The resulting FPE is exact and correctly reduces to that of Ref. \cite{Adelman1} in the classical Markovian limit. The absence of the initial preparation of the system in the Feynman-Vernon theory was incorporated by Grabert et al. \cite{GrabertSchrammIngold} in their functional integral formalism of quantum dissipation. It validated the findings already derived in Ref. \cite{SchrammJungGrabert}. Subsequently, Allahverdyan et al. \cite{AllahverdyanNieuwenhuizen} propose an alternate FPE in their study about the extraction of the work of a quantum particle strongly coupled to the bath. The authors \cite{NieuwenhuizenAllahverdyan} fully derived it later on in their proof that at low temperatures Clausius inequality is violated \textcolor{black}{due to the appearance of quantum coherence by the off-equilibrium nature of the bath. That is, a construction of  a perpetual model of the second kind.
This assertion has generated a debate between coherence and entanglement as the cause of the violation. Very recently, Soltanmanesch et al. \cite{SoltanmaneshShafiee} experimentally validated it using interferometry techniques while Micadei et al. \cite{MicadeiEtAl} shown by means of Nuclear Magnetic Resonance, that two initially thermal correlated spin-1/2 break the time-reversal symmetry and changes the direction of the energy flow between the qbits. For a comprehensible theoretical analysis about quantum coherence and quantum correlation see Fan et al. \cite{FanPengEtAl} and the references therein.} 

\textcolor{black}{References [3] - [11] dated back almost 20 years ago or more.} In all these works, the second derivatives of the FPE for the Wigner function \textcolor{black}{$W(p,q,t)$} in the whole phase space  are given in terms of the different diffusion functions of the system. It should be said that the generalized diffusion has not been calculated in these works.

\textcolor{black}{ A more recent search for related researches yields that the detailed article of 2011 by Fleming et al. \cite {FlemingRouraHu}, the 2017 one by Carlesso et al. \cite {CarlessoBassi} and that of 2018 by Shen et al. \cite {ShenSuZhouYi}, to name a few, shows numerical results for the standard deviation and its relation to the generalized diffusion. Their results along with those of this investigation will be analyzed later in Sec. \ref{secB}. }

Recently, a rather simple method  \cite{PJPRE} based on Ref. \cite{SchrammJungGrabert} was developed to analyze the system in the configuration space. There,  the FPE of the reduced Wigner function \textcolor{black}{$W(q,t)$}, its standard deviation and the generalized diffusion are properly defined in terms of the temperature, the friction damping and the characteristic quantum frequency of the bath.     

This article will use \textcolor{black}{the results of Ref. \cite{PJPRE} to provide a methodology to calculate} the standard deviation associated to the reduced Wigner function \textcolor{black}{in the configuration space}, the diffusion function appearing in its FPE and their comparison with those of the continuum limit. \textcolor{black}{It differs from previous calculations because of the inclusion  in the description of the system initial quantum preparation. It will be shown that the generalized diffusion is negative for specific values of the friction coefficient of the Ohmic thermal bath as the already predicted \cite{FlemingRouraHu}. However, it shows new features because of the initial system preparation. It is mathematically consistent with the physics of the problem having a simple explanation in terms of an interplay of the susceptibilty of the system to the external field and, the quantum entanglement of the particle position with the reservoir quantum noise.} The classical time-dependent diffusion term is directly obtained from the constitutive equations of the \textcolor{black}{quantum} scenario. It also shows a similar behavior although in a lesser magnitude.

The manuscript is structured as follows. In Sec. \ref{secA} is summarized the theoretical framework of this proposal. The methodology used in the calculations and the discussion of the numerical results is presented in Sec. \ref{secB}. The manuscript concludes in Sec. \ref{secC} with a set of general remarks and a prospective for future research.

\section{Relevant equations}
\label{secA}
For the purpose of a better comprehension, a brief account of the main equations is given below \cite{PJPRE}. This will allow to identify the different terms contributing to the final expressions of the diffusion term and the standard deviation of the \textcolor{black}{reduced} Wiener function $W(q,t)$. The latter is also the conditional probability $p(q,t|\qzero)$ to find the particle at $\{q,t\}$ given that at initially was in $\qzero$. 

The dynamics for the position $q(t)$ of a particle with mass $M$, subjected to an external potential $\omega_{0}^{2}q^{2}(t)/2$, in contact with an Ohmic bath of quantum harmonic oscillators at  temperature $T$ with a friction coefficient $\gamma$, is given by the following \textcolor{black}{quantum} Langevin equation (\textcolor{black}{QLE}):
\be
\ddot{q}(t)=-\gamma\,\dot{q}(t)-\frac{\omega_{0}^{2}}{M}\,q(t)+\frac{1}{M}\,\xi(t),\\
\label{QLE}
\ee
where the quantum noise $\xi(t)$ is characterized by a zero mean and correlation functions:
\bse
\bea
\left<\xi(t-s)\,\xi_{0}\right>=&-&\left(\frac{\gamma M}{2\beta}\right)\nu\sinh\left[\frac{1}{2}\nu(t-s)\right]^{-2}\nonumber\\
&+&\,i\,\gamma\, M\,\hbar\,\dot{\delta}(t-s),\label{corxi}\\
\left<\xi(t)\qzero\right>&=&-\frac{2\,\gamma}{\beta}\sum_{\mathbf{n}=1}^{\infty}\frac{\nun}{(\nun\!+\!\lambdaone)(\nun\!+\!\lambdatwo)}\mbox{e}^{-\nun t}.\label{xiq0}
\eea
\ese
Here, $\hbar$ is the Planck constant divided by $2\pi$, $\beta=(\kb T)^{-1}$, $\kb$ is the Boltzmann constant, $\nun=n\nu$ with $\nu=2\pi(\hbar\beta)^{-1}$ and $ \lambda_{_{\mathbf{1,2}}}=(\gamma\pm(\gamma^{2}- 4\omega_{0}^{2}/M)^{1/2}\,)/2$.

\textcolor{black}{The probability distribution function associated to the QLE is the Wigner function $W(p,q,t)$ which in turn is the representation in phase space of the full quantum master equation for the density matrix operator \cite{SchrammJungGrabert}.}

The Laplace transformation of Eq. (\ref{QLE}) gives the result: 
\bea
\dot{q}(t)&=&\overline{v}(t)+\phiv(t),\label{qprime}
\eea
where $\overline{v}(t)=\chiqp(t)\qzero+\chivp(t)\vzero$ is the mean drift velocity with $\vzero$ being its initial value. The noise $\phiv(t)$ is a functional of $\xi(t)$ and the susceptibilities $\chiq(t)$ and $\chiv(t)$  are, respectively:
\bse
\bea
\phiv(t)&=&\frac{1}{M}\int_{0}^{t}ds\chivp(t-s)\xi(s),\label{varphiv}\\
\chiq(t)&=&\mbox{e}^{-\gamma t/2}\Bigg(\!\cosh\left[\frac{\omega\,t}{2}\right]\!+\!\frac{\gamma}{\omega}\sinh\left[\frac{\omega\,t}{2}\!\right]\!\Bigg),\label{chiqs}\\
\chiv(t)&=&\frac{2}{\omega}\mbox{e}^{-\gamma t/2}\sinh\left[\frac{\omega\,t}{2}\right]\label{chivs},\\
\omega^{2}&=&\gamma^{2}-\frac{4\omega_{0}^{2}}{M}.
\eea
\ese

\textcolor{black}{The FPE associated to the probability distribution in the configuration space $p(q,t)$, which is also the reduced Wigner function $W(q,t)$, is given by \cite{PJPRE}:}
\bse\bea
\frac{\partial p(q,t)}{\partial t} &=&-\Omega(t)
\frac{\partial}{\partial q}\Big[q\,p(q,t)\Big]+\frac{1}{2}\DQ(t)\frac{\partial^{2}p(q,t)}{\partial q^{2}},\label{adelman1}\\
\Omega(t)&=&\frac{\chiqp(t)}{\chiq(t)}=\frac{d\ln\chiq(t)}{dt},\label{Omega1}\\
\DQ(t)&=&\sigmaQp(t)-2\,\sigmaQ(t)\,\Omega(t)\label{Dadelman},\\
\sigmaQ(t)&=&\sigmaone\!(t)+\frac{\kb T}{M}\chiv^{\!\!2}(t).\label{SigmaAveraged}\\
\sigmaone\!(t)&=&\int_{0}^{t}\!\!d\tp \Done(\tp)\label{A11},\\
\Done(t)=&2&\Bigg[\!\int_{0}^{t}\Big<\phiv(t)\phiv(\tp)\Big>d\tp\nonumber\\
&+&\chiq(t)\Big<\phiv(t)q(0)\Big>\Bigg].\label{D1clas}
 \eea\ese

Function $\Omega(t)$ is a hydrodynamic-like drift frequency and $\DQ(t)$ and $\sigmaQ(t)$ are the \textcolor{black}{generalized} diffusion of the bath and the standard deviation of the reduced Wigner function, respectively.

\subsection{Noise correlation functions}
\label{secA1}
From now on, all quantities will be expressed by scaling energy, position and time by the parameters $(\omega_{0}^{2}\qzero^{2})$, $\qzero$ and $M^{1/2}\omega_{0}^{-1}$, respectively.

As is is seen from Eqs. (\ref{Dadelman}) and (\ref{SigmaAveraged}), $\sigmaQ(t)$ and $\DQ(t)$ depend on the correlation functions $\left<\phiv(t-s)\,\phivzero\right>$ and $\left<\phiv(t)\,\qzero\right>$, respectively. In particular.
\bea
\left<\phiv(t-s)\,\phivzero\right>&=&\int_{0}^{t}\chivp(t-x)\,dx\int_{0}^{s}\chivp(s-y)\nonumber\\
&\times&\left<\xi(x-y)\,\xi_{0}\right>\,dy.
\eea

 Then, having used Eq. (\ref{corxi}) it is found,
\bea
\big<\phiv(t)&&\hspace{-11pt}\phivzero\big>=-\gamma\, T\int_{0}^{t}\chivp(t-x)\,dx\int_{0}^{s}\chivp(s-y)\nonumber\\
&\times&\Big[\frac{\nu}{2}\csch^{2}(x-y) +i\,\frac{2\pi}{\nu}\,\partial_{x}\delta(x-y)\Big]dy
\label{noiseintegral}
\eea
This is a complicated integral to solve due to the singularity shown by $\csch(x-y)$ at $x=y$. To bypass this inconvenience, it is appealed  to Eq. (\ref{qprime}), from which,
\begin{widetext}
\bea
\big<\phiv(t-s)\,\phivzero\big>&=& \Big<v(t-s)v_{_{0}}\Big> -\chivp(t)\Big<v(s)\vzero\Big>
-\chivp(s)\Big<v(t)\vzero\Big>-\chiqp(t)\Big<v(s)\qzero\Big>
-\chiqp(s)\Big<v(t)\qzero\Big> +\chivp(t)\chivp(s)\Big<\vzero^{2}\Big>\nonumber\\
&+&\Bigg[\chivp(t)\chiqp(s) + \chivp(s)\chiqp(t)\Bigg]\Big<\vzero\qzero\Big>
+\chiqp(t)\chiqp(s)\Big<\qzero^{2}\Big>.\label{phitphis}
\eea
\end{widetext}
These correlations are easily obtained in terms of the position correlation function, i.e \cite{GrabertWeissTalkner}
\bse
\bea
\Big<v(t-s)\vzero\Big>&=&-\frac{\partial^{2}}{\partial t^{2}}\Big<q(t)\qzero\Big>\Big|_{t=t-s},\\
\Big<v(t)\vzero\Big>&=&-\frac{\partial^{2}}{\partial t^{2}}\Big<q(t)\qzero\Big>,\\
\Big<v(t)\qzero\Big>&=&\frac{\partial}{\partial t}\Big<q(t)\qzero\Big>,\\
\Big<\vzero\qzero\Big>&=&\frac{\partial}{\partial t}\Big<q(t)\qzero\Big>\Big|_{t=0},
\eea
\ese
where $\Big<q(t)\qzero\Big>$ is given by \cite{GrabertWeissTalkner,Ingold}:
\bse
\bea
\Big<q(t)\qzero\Big>&=&S(t)+i A(t),\label{corpos}\\
S(t)&=&\frac{\pi}{2\,\gamma\,\omega}T\Bigg[\cot\big[\frac{\pi\lambda_{2}}{\nu}\Big]\mbox{e}^{-\lambda_{2}t}\nonumber\\
&-&\cot\big[\frac{\pi\lambda_{1}}{\nu}\Bigg]\mbox{e}^{-\lambda_{1}t}-\Gamma(t),\\
A(t)&=&-\frac{\pi\, T}{\omega\,\nu}\sinh\big[\omega\, t\big]\mbox{e}^{-\gamma t/2},\\
\Gamma(t)&=&2\,\gamma\,T\sum_{\mathbf{n}=1}^{\infty}\frac{\nun}{(\nun\!+\!\lambdaone)(\nun\!+\!\lambdatwo)},
\eea
 \ese
where \textcolor{black}{$\lambda_{1,2}=(\gamma\pm\omega)/2$}. The  first term of Eq. (\ref{corpos}) describes how the particle is driven away from equilibrium by the external potential while the second, the relaxation when the random force holding it away from equilibrium is released. 


The dispersion $\big<\vzero^{2}\big>$ diverges logarithmically because the Ohmic system depends on a frequency-independent damping with a non-vanishing value at infinite. Considering a thermal bath with a high frequency cutoff as in real systems, this  inconvenience is solved. It is the so called Drude regularization in which the damping has a finite memory $\omegaD$, the Drude frequency, so that at very short times of order $1/\omegaD$  the system behaves Ohmic. This line of reasoning leads to that the dispersions needed in Eq. (\ref{phitphis}) are  \cite{GrabertWeissTalkner,GrabertSchrammIngold}
\bse
\bea
\Big<\qzero^{2}\Big>&=&2\,T\sum_{n=1}^{\infty}\frac{1}{1+\nun^{2}+\gamma\,\nun},\\
\Big<\vzero^{2}\Big>&=&2\,T\sum_{n=1}^{\infty}\frac{\omegaD+\nun+\gamma\,\omegaD\,\nun}{(1+\nun^{2})(\omegaD+\nun)+\gamma\,\omegaD\,\nun}.
\eea 
\ese

To complete the set of correlation functions necessary for the calculation, the second correlation of Eq. (\ref{D1clas}) still needs to be defined. This is given by:
\bea
\big<\phiv(t)\qzero\big>&=&-2\,\gamma\,T\int_{0}^{t}\chivp(t-x)\big<\xi(x)\qzero\big>\, dx.
\label{phivq0}
\eea

As before, this equation can be rewritten by using Eq. (\ref{qprime}), i.e
\be
\big<\phiv(t)\qzero\big>=\big<v(t)\qzero\big> - \chivp(t)\big<\vzero\qzero\big>-\chiqp(t)\big<\qzero^{2}\big>,
\ee
where
\be
\big<\vzero\qzero\big>=\big<v(t)\qzero\big>\Big|_{t=0}.
\ee

In summary, the diffusion $\DQ(t)$ and the standard deviations $\sigmaQ(t)$  are easily calculated as derivatives
of the correlation function $\big<q(t)\qzero\big>$ instead of solving the multidimensional integral given by Eq. (\ref{noiseintegral})
and Eq. (\ref{phivq0}), respectively. They are only given in terms of the friction damping, the temperature and the quantum time scale $1/\nu$.
\subsection{Continuum limit}
\label{secA2}

The classical Markovian case is obtained by setting $\left<\xi(t-s)\,\xi_{0}\right>=2\,\gamma\, T\, \delta(t-s)$. In this scenario, the correlation $\left<\phiv(t-s)\,\phivzero\right>$ is \footnote{Mathematica\texttrademark\, computer package v 6.0.2.0 was used for the algebraic, numerical and graphics manipulations.}:

\bea
\big<&\phiv&(t\!-\!s)\phivzero\big>_{\mbox{\tiny{\!\bf{clas}}}}=\frac{T}{2\omega^{2}}\mbox{e}^{-(\gamma+\omega)(t+s)}\Bigg[\gamma^{2}(1-\mbox{e}^{\omega s})\nonumber\\
&\times&(\mbox{e}^{\omega t}-1)+\gamma\,\omega\bigg(\mbox{e}^{\omega(t+s)}+\mbox{e}^{(\gamma+\omega)s}-\mbox{e}^{\omega t+\gamma s}-1\bigg)\nonumber\\
&+&\omega^{2}\big(\mbox{e}^{\omega s}-1\big)\big(\mbox{e}^{\omega t}+\mbox{e}^{\omega s}\big)\Bigg],
\label{MarkovNoise}
\eea
from which the diffusion term and the standard deviation according to Eqs. (\ref{Dadelman}) and (\ref{SigmaAveraged}) are, respectively
\bse
\bea
D_{\mbox{\tiny{\!\bf{clas}}}}(t)&=&\frac{4\,T}{\gamma+\omega\,\coth\big[\frac{\omega}{2}\,t\big]},\label{MarkovDif}\\
\sigmac &=&T\Bigg[1-\frac{\mbox{e}^{-\gamma\,t}}{\omega^{2}}\Big(\big(\gamma^{2}-2\big)\cosh\big[\omega t\big]\nonumber\\
&+&\gamma\,\omega\sinh [\omega\,t]-2\Big)\Bigg].\label{MarkovSigma}
\eea
\ese

The FPE given by Eq. (\ref{adelman1})  is identical to the one obtained by  Adelman et al. \cite{AdelmanGarrison} for the classical generalized Langevin equation if the quantum contribution is dismissed. It differs from the phenomenological equation of Chandrasekhar \cite{Chandrasekhar} in that the diffusion coefficient is not a constant but a generalized function of time.
The diffusion term given by Eq. (\ref{Dadelman}) has not been  explicitly developed in terms of the parameters of the model. Up to the knowledge of the author, Eq. (\ref{MarkovDif}) has not been reported in the literature although it is a natural result coming from the dynamics. It will allow to examine the dependence of the classical generalized diffusion with the frequency $\omega$. 
Equation (\ref{MarkovSigma}) reduces to the classical result by adding the term $T\chiv^{\!\!2}(t)$ to the standard deviation appearing in Eq. (213) of Ref. \cite{Chandrasekhar}.

\section{Discussion of the results}
\label{secB}

Unfortunately there is no experimental data to compare with the \textcolor{black}{quantum} description. However, the order of magnitude of the friction coefficient at temperatures where quantum effects are important can be approximately inferred from experiments at room temperature. Thus, Zchang \cite{Zhang} found that for polystyrene spherical particles of radius $0.5\, \mathrm{\mu m}$ and mass of  $54.9$ pKg \footnote{Information of some physical properties of  polystyrene solutions can be obtained from:\newline http://www.polysciences.com/skin/frontend/default/\newline polysciences/pdf/TDS\%20238.pdf}  immersed in water (viscosity $\eta=1.002\, \mathrm{\mu m}^{2}\mbox{s}^{-1}$)  at $293.15\,\mathrm{K}$  trapped in an optical tweezer, the stiffness of the trap is $\omega_{0}^{2}=1.6\,\mathrm{\mu Kg\,s}^{-2}$. Considering  this values gives a friction damping $\gamma$ of $17.17\, \mbox{MHz}$ and $\omega_{0}\,M^{-1/2}=53.98\,\mbox{KHz}$. Using the scale parameters and assuming the bead is initially located $100\,\mu\mbox{m}$ off the center of the potential, the reduced temperature is of $25.28$  and the reduced friction coefficient is 318.06, respectively.  This value of $\gamma$  says that the classical dynamics is highly overdamped.

As it was shown above, all calculations in the \textcolor{black}{quantum} approach depend on the position correlation function. Their outcomes depend on the value of the frequency $\omega=\sqrt{\gamma^{2}-4}$ from which three responses of the dynamics to the value of $\gamma$ can be found. Namely,  periodic ($\gamma<2$; $\omega$ imaginary), aperiodic ($\gamma=2$; $\omega=0$) and overdamped ($\gamma>2, \omega$ real). It will be arbitrarily supposed that the stiffness of the trap has a value of $18.2\, \mathrm{mKg\,s}^{-2}$ and the temperature is 70 K. For the same $M$ and $\qzero$ used before, $\gamma$ values of 1, 2 and 4 requires the bath should has hypothetical friction coefficients of 5.75, 11.5 and 53.98 GHz, respectively. The reduced temperature will be 0.053 and frequency $\nu=10^{7}$. This choice of parameters will allow to compare the dynamics around the aperiodic behavior.

  \begin{figure}[h]
 \begin{center}
\includegraphics[height=11cm,width=9.5cm,angle=0]{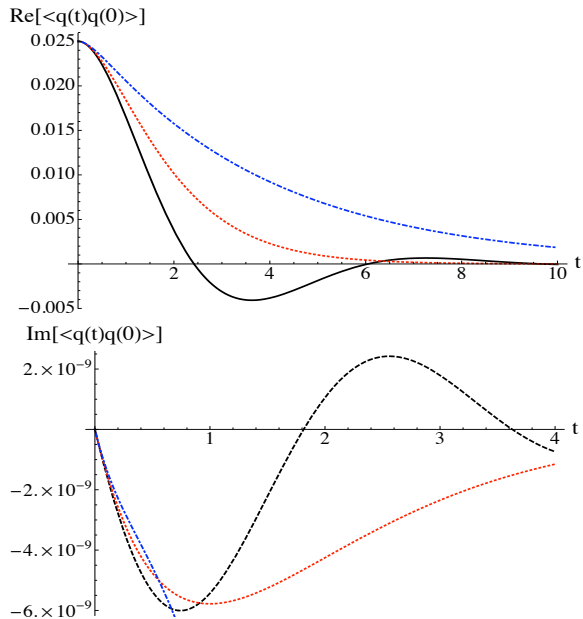}
\vspace{-2.75cm}
\caption{Real and imaginary parts of the position correlation function $\big<q(t)q(0)\big>$ versus time $t$ for $\gamma$ values of 1.0 (solid black), 2.0 (dotted red) and 4.0 (dot-dashed blue); $T=0.053$ and $\nu = 10^{7}$.}
\label{fig1}
\end{center}
\end{figure}

This is exemplified in Fig. \ref{fig1} where the position correlation function $\big<q(t)\qzero\big>$ is plotted for  $\gamma$ values of 1 (solid black), 2 (dotted red) and 4 (dot-dashed blue), respectively. 
It can be seen in the two panels of the figure that $\gamma=2$ defines a separation domain where the oscillatory behavior of the correlation function begins. Even though the imaginary contributions (bottom panel) are negligible in comparison with the real counterparts, at large times (not shown) the overdamped case (dot-dashed blue curve) becomes much smaller than the others two. These features do not produce a significant change in $ \sigmaQ (t) $, except in their magnitudes. However, the diffusion term in the imaginary $\omega$ scenario will be very different. This is discussed next.

The statistics of random complex numbers require that the standard deviation, in addition to being positive, must also be equal to the sum of the standard deviations of real and imaginary parts. It means that the calculations should be done on each component of the position correlation function so that $\sigmaQ(t)$ is the sum of the two contributions. The diffusion term $\DQ(t)$ follows according to Eq. (\ref{Dadelman}). The real part of these two functions is concerned with the spreading of the probability
density due to the external potential and interference effects due to the quantum coherence, while the imaginary part to the dissipation by the quantum thermal bath.

\begin{figure}[h]
 \begin{center}
\includegraphics[height=11cm,width=9.5cm,angle=0]{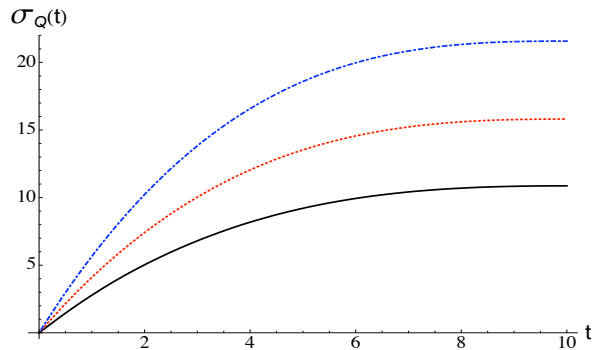}
\vspace{-6.3cm}
\caption{Standard deviations $\sigmaQ(t)$ versus $t$ for the curves described in Fig. (\ref{fig1}).}
\label{fig2}
\end{center}
\end{figure}

Figure \ref{fig2} shows the standard deviation up to $t=10$. The time dependence of the real and imaginary contributions are always positive. From the graph it seems that $\sigmaQ(t)$ reaches a maximum, however it is a continuos growing function of time. It means that the initial equilibrium distribution is widen with the position as time progresses so at very large values of $t$ there is no a preferable position for the particle to be found. This is a consequence of the quantum entanglement with the heat bath.  Frequency $\omega$  changes the magnitude of the function with no visible effect wether or not $\omega$ be real or imaginary. The significance of $\gamma<2$ in the dynamics is better viewed by examining the time-dependent diffusion. 

It is generally accepted that equilibrium classical negative diffusion coefficients would denote process of "concentration" as opposed to diffusion. Thus, entropy is decreasing. That means, it's not a random walk problem but additional forces are acting opposing to diffusion process resulting in concentration. This paradigmatic vision of diffusion is not fulfilled in the periodic \textcolor{black}{quantum} case, even in the classical one, as it will be demonstrated next.

A system at equilibrium under the effect of an external force is prohibited to have negative diffusion because of the Stokes-Einstein relationship $D=\kb T/\gamma$. The particle cannot move against the field. However for a non-equilibrium situation there is not any fundamental principle that disallows it to occur. This is what was found in several classical problems such as in the coupled phase response of an ensemble of oscillators \cite{ReimannVanDerBroeckKawai}, overdamped Brownian motion through a ``corridor'' with periodic and symmetric obstacles \cite{EichhornReinmannHanggi}, random walk  in which the step directions taken by the walker are correlated with previous ones \cite{CleurenVanDerBroeck2002} and Brownian motion biased by a potential in a layered domain with unbiased transitions between \cite{CleurenVanDerBroeck2003}. Although, according to the authors, those specific dynamical models can mimic specific Brownian motion problems, unfortunately, negative diffusion has so far not been experimentally observed. \textcolor{black}{However, for the quantum Brownian motion it was found in 2011 by Fleming et al. \cite{FlemingRouraHu} that for a given set of model parameters and initially particle states uncorrelated with the environment, the diagonal position element of the covariance matrix  accompanying $W(p,q,t)$ oscillates as time progresses. Using the relationship of it with the diffusion term, it was predicted that it shows negative values. This  feature seems to indicate that a generalized diffusion showing negative values is a trait own to the quantum Brownian motion. However, as it will be proved later, the negativeness even persists in the classical regime for a definite set of parameters. Although the 2017's Carlesso et al. research  \cite{CarlessoBassi} worked out the same problem for an overdamped regime in the Heisenberg picture, it would not be discarded that a negative diffusion would be found for an underdamped bath. Very recently,  in 2018 Shen et al. \cite{ShenSuZhouYi} showed that for an external force given by an electric field and uncorrelated initial preparation, the time-dependent coefficient related to the generalized diffusion of the master equation under the Rotating Wave Approximation also shows a fluctuating covariance matrix. The main difference of these works with the one presented in this research is the inclusion  in the latter of the correlation between the bath noise and the initial particle position. In most of physical situations such correlations are present and therefore they have to be taken into account.  In a work due to Karrlein et al. \cite{KarrleinGrabert} it was  shown that switching the interaction at $t=0$, uncorrelated system, affects the dynamics at long times: the classical limit of Adelman et al. \cite{AdelmanGarrison} is not recovered and the Fokker-Plank operator terms differ with those of the correlated. Additionally the work by \v{S}telmachovi\v{c} et al. \cite{StelmachovicBuzek} showed that the parameters describing an uncorrelated system--bath entanglement ``cannot be determined by performing a local measurement on the initial state of the system'', that is to say, on the Brownian particle. Fonseca Romero et al. \cite{FonsecaTalknerHanggi} illustrate that for several examples of open quantum systems. }

\textcolor{black}{Under the premise of existence of an initial particle-bath entanglement, the results reported here describe in more detail the physical picture of the problem under consideration.}

\begin{figure}[h]
 \begin{center}
\includegraphics[height=14cm,width=12.cm,angle=0]{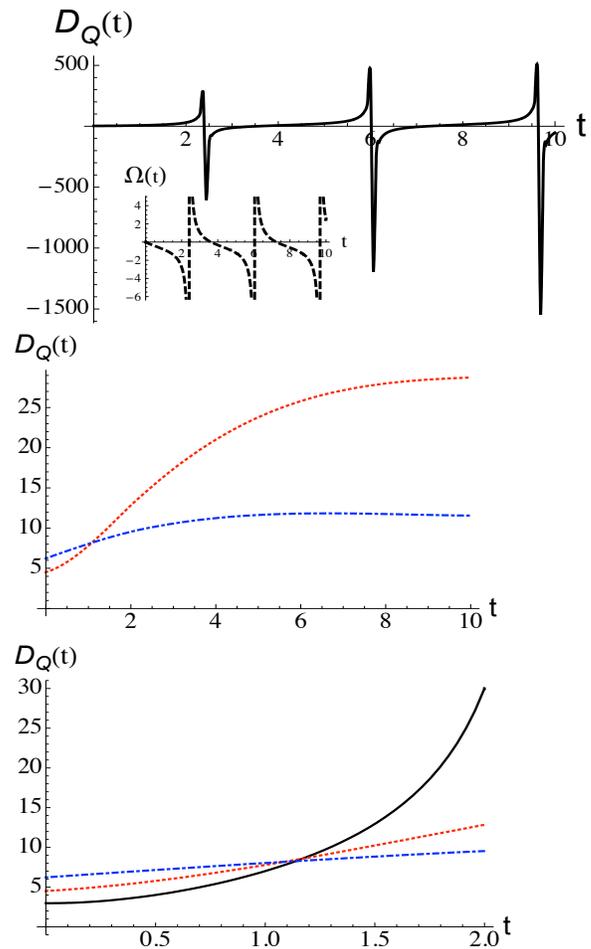}
\vspace{-1.5cm}
\caption{Quantum generalized diffusion $\DQ(t)$ versus $t$ for the curves described in Fig. (\ref{fig1}). The inset in the top panel shows the time dependence of $\Omega(t)$ for $\gamma=1$. The bottom panel is an enlarge at short times.}
\label{fig3}
\end{center}
\end{figure}

This is revealed in the top  of Fig. \ref{fig3} for $\gamma=1$. The middle panel shows the function for
$\gamma$ of 2 (dotted red) and 4 (dot-dashed blue), while  the bottom panel has short scale to show the value $\DQ(0)$.

The top panel shows negative values of $\DQ(t)$ at some time intervals and strong discontinuities, as well. Besides the quantum contribution there is an additional one coming from the susceptibility $\chiq(t)$ in the $\Omega(t)$ function given by Eq. (\ref{Omega1}). It is shown in the inset of the figure for $\gamma=1$. It does not approach to a long--time limit and is a very discontinuing function for $\gamma<2$  with poles and zeroes along the real time axis. The quantum entanglement from the standard deviation modifies its magnitude while $\Omega(t)$ provides the discontinuous oscillatory character to $\DQ(t)$ and its brief partial negativity for short time lapses as well. The quantum contribution is due to the effect of the cloud of the thermal bath oscillators surrounding the particle. It generates a relevant role in the dynamics, particularly at low temperatures. This nontrivial effect was pointed out by Nieuwenhuizen et al. \cite{NieuwenhuizenAllahverdyan} in the derivation of the full space Wigner function $W(x,\mbox{p},t)$ for the generalized quantum Brownian motion. In any case, this ``strange'' dynamical response of $\DQ(t)$ is mainly due to the contribution of $\Omega(t)$. This feature in the periodic regime \textcolor{black}{is the main result of this work. It will be shown later that it is also be observed in classical systems.}

Then, it should be said that the periodic result found in this proposal is own to the physics of a non-equilibrium stationary system due to the absence of a temporal finite limit in $\DQ(t)$.

There is another striking feature in the \textcolor{black}{quantum} approach shown in the bottom panel of Fig. \ref{fig3}. It concerns with the initial value of $\DQ(t)$, which is not zero as expected. It is primordially due to the initial preparation of the system. In other words, to the correlation of the quantum noise with the initial position of the Brownian bead.

 \begin{figure}[ht]
 \begin{center}
 \includegraphics[height=15cm,width=12.cm]{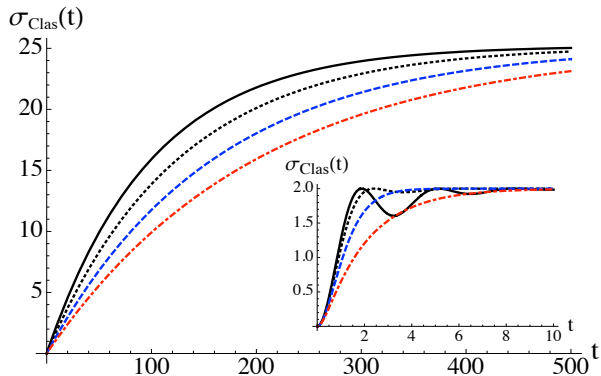}
\vspace{-9.5cm}
\caption{Classical standard deviations $\sigmac(t)$ versus $t$ for $\gamma$ values of 200 (solid black), 250 (dotted black), 318 (dashed blue) and 400 (dot-dashed red); $T=25.2$. The inset corresponds to $\gamma$ of 0.5 (solid black), 1.0 (dotted black), 2 (dotted blue) and 4 (dot-dashed red) at $T=2$. }
\label{fig4}
\end{center}
\end{figure}

The classical counterpart for $\sigmaQ(t)$ is shown in Fig. \ref{fig4} for $\gamma$ lying around 318 corresponding to the overdamped system mentioned in the beginning of this section. No special characteristics are observed except the value predicted by the equipartition principle at very large values of $t$. However, this function is sensitive to the value of $\omega$. For a hypothetical classical system at $T=2$ and $\gamma$ values of 0.5, 1, 2 and 4, the inset of Fig. \ref{fig4} shows oscillations in the periodic regime at specific times.

\begin{figure}[ht]
 \begin{center}
\includegraphics[height=11cm,width=9.cm,angle=0]{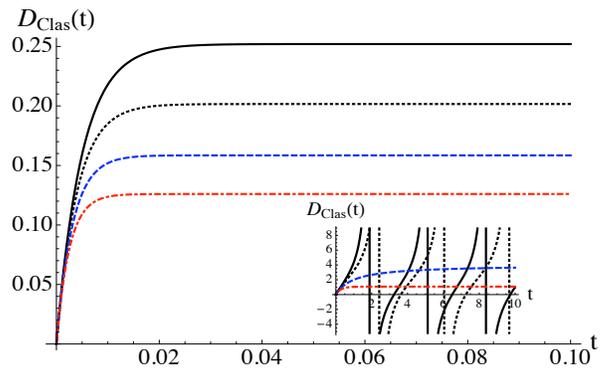}
\vspace{-6.25cm}
\caption{Classical diffusion $D_{\mbox{\tiny{\!\bf{clas}}}}(t)$ versus $t$ for the curves described in Fig. (\ref{fig4}).}
\label{fig5}
\end{center}
\end{figure}

The associated $D_{\mbox{\tiny{\!\bf{clas}}}}(t)$ is presented in Fig. \ref{fig5}. The inset shows the corresponding effect of $\Omega(t)$ for imaginary $\omega$. It has similarities with the \textcolor{black}{quantum} and tells that the Markovian classical Brownian motion will also predict transient negative diffusion during times lapses of the dynamics if and only if it is periodic, that is, $\gamma$ is less than 2. It has physical consistency, although the discontinuity and negativeness of $D(t) $ might seem `` un-physical '' in the paradigmatic context of the problems of equilibrium diffusion and  concentration processes mentioned above.
 
In summary, results for an imaginary $\omega$ could probably be achieved in the \textcolor{black}{quantum} scenario due to the requirement of low temperatures to reach that particular dynamic state. The relevance of the theoretical prediction on both regimes depend on its experimental confirmation.

\section{Concluding remarks}
\label{secC}

The results for the calculation of the generalized diffusion $\DQ(t)$ of a Brownian particle interacting with a bath of quantum harmonic oscillators have been presented. They are anchored on strong  mathematical and physical grounds.

It was found that the bath spread out the initial equilibrium probability and modifies  the positiveness of the diffusion term during brief period of time when the system behaves periodic. The nature of it is that the dynamics is own of a stationary non-equilibrium system with a strong contribution coming from the periodic and discontinued frequency $\Omega(t)$. 

Additionally, the initial  behavior of $\DQ(t)$ is finite in contrast to the classical result. It is ascribed to the correlation of the quantum noise and the initial position of the particle. At any time and $\omega$ value, the entanglement of the bath with the particle will substantially modified the dynamics in comparison with the classical one.


As it was mentioned in Ref. \cite{PJPRE}, the \textcolor{black}{quantum} description reduces correctly to classic when the quantum noise is white. 

The negative aspect of the generalized diffusion function in periodic systems went unnoticed in the classical bibliography of the Brownian motion, despite the fact that Adelman et al. \cite{AdelmanGarrison} made an analysis of it. In general, the standard deviation, mean square displacement, the Gaussian probability distribution and the long time diffusion limit, to cite a few, were among the main properties to determine, mainly in the overdamped regime or high friction limit. No mention to the characteristics of the generalized diffusion in the underdamped or periodic scenario has so far been documented. 

The results obtained are mathematically consistent and their validity will depend on the experimental feasibility of building a quantum system under these conditions. 

Finally, there could be a potential extension of this investigation. It has to do with the  similarities shared by Eq. (\ref{adelman1}) with the FPE of an Ornstein-Uhlenbeck process (OUP) with variable coefficients \cite{Risken,Gardiner1}. 
In fact, by means of Ito's rule \cite{Gardiner1}, the \textcolor{black}{c-number} dynamics can also be interpreted as the following first-order stochastic differential equation,
\be
\dot{q}(t)=-\Omega(t)\,q(t) +\sqrt{\DQ(t)}\,\zeta(t),
\label{OUP}
\ee 
where $\zeta(t)$ is a Gaussian zero mean delta correlated white noise. Its solution provides the variance of its Gaussian probability distribution from which $\DQ(t)$ is calculated by Eq. (\ref{Dadelman}).

In this respect, the dynamics of the \textcolor{black}{QLE} is simplified to an OUP, Eq. (\ref{OUP}), with the ubiquitous white noise term. 
The quantum contribution is propagated by the time-dependent diffusion function $\DQ(t)$ and the response of the system to the external field by the frequency $\Omega(t)$. 
Since it is well known the applicability of this stochastic scheme to classical problems as anomalous diffusion, exit and first passage times and dynamic relaxation, to name a few, then it potentially could be extended to the \textcolor{black}{quantum} scenario. 
It is not a methodology to replace it, but instead,  it is an excellent theory to analyze the quantum contribution to some other  problems of physical interest. The only requirement is that generalized diffusion $\DQ(t)$ and frequency $\Omega(t)$ must have been first determined from the solution of the \textcolor{black}{QLE} in the configuration space. 

\begin{acknowledgments}

This work has profited from correspondences with Dr. Oscar Paredes-Altuve from Universidad de Chile.
\end{acknowledgments}

\bibliographystyle{apsrev4-1}
\bibliography{references}
\end{document}